\documentclass[11pt]{gh2013}

\usepackage{mathptmx}       
\usepackage{helvet}         
\usepackage{courier}        
\usepackage{makeidx}         
\usepackage{graphicx}        
\usepackage{multicol}        

\begin{document}

\title*{Super Star Clusters in Luminous Infrared Galaxies: the SUNBIRD Survey}
\titlerunning{Super Star Clusters in LIRGs}
\author{P. V\"ais\"anen$^{1,2}$, Z. Randriamanakoto$^{1,3}$, A. Escala$^{4}$, E. Kankare$^{5}$, A. Kniazev$^{1,2}$, J.K. Kotilainen$^{5}$, S. Mattila$^{5}$, R. Ramphul$^{1,3}$, S. Ryder$^{6}$, A. Tekola$^{1,7}$}
\authorrunning{Vaisanen et al.} 
\institute{$^{1}$ South African Astronomical Observatory, P.O. Box 9 Observatory, Cape Town, South Africa \email{petri@saao.ac.za}
\\ $^{2}$ Southern African Large Telescope, P.O. Box 9 Observatory, Cape Town, South Africa
\\ $^{3}$ University of Cape Town, Astronomy Department, Private Bag X3, Rondebosch 7701, South Africa
\\ $^{4}$ Departamento de Astronom\'ia, Universidad de Chile, Casilla 36-D, Santiago, Chile
\\ $^{5}$ Finnish Centre for Astronomy with ESO (FINCA), University of Turku, V\"ais\"al\"antie 20, FI-21500 Piikki\"o, Finland
\\ $^{6}$ Australian Astronomical Observatory, P.O. Box 915, North Ryde, NSW 1670, Australia
\\$^{7}$ Las Cumbres Observatory Global Telescope Network, Goleta, CA, 93117, USA}%
%
\maketitle


\vskip -3.0 cm  
\abstract{
We present recent results from an adaptive optics imaging survey of 40 Luminous IR Galaxies (LIRGs) 
searching for obscured core collapse supernovae and studying the galaxies themselves.  Here, in particular, 
we discuss the Super Star Clusters (SSC) populations in the LIRGs.  We have constructed the
first statistically significant samples of Luminosity Functions (LF) of SSCs in the
near-IR, and find evidence that the LF slopes in LIRGs are shallower than in 
more quiescent spiral galaxies.  Distance and blending effects were investigated
in detail paving the way for SSC studies further out than done previously.  
We have also correlated the luminosities of the brightest clusters with the star formation rates (SFR) of 
the hosts. The relation is similar, though somewhat steeper than that found in the optical and at 
lower SFR levels, suggesting systematic extinction and/or age effects. We find that the 
characteristics of the relation suggest an underlying physical driver rather than solely a size-of-sample
effect.  In particular, a truncated luminosity/mass function would naturally explain the small scatter
we find. Finally, we are modelling the ages and masses of our near-IR
detected clusters in conjunction with HST optical data and present early 
results of using SSC properties to trace the histories of the target LIRG systems.
}

\section{Introduction: The SUNBIRD Survey}
\label{sec:intro}

We are conducting a survey of approximately 40 Luminous Infrared Galaxies (LIRGs; galaxies with $10^{11} >  L_{IR}  > 10^{12}$)  using adaptive optics (AO) NIR imaging  with VLT/NACO, Gemini/ALTAIR/NIRI, and recently also using the superb "wide field" capability of multi-conjugate AO on Gemini/GEMS.  The instruments deliver images with a spatial resolution perfectly complementing existing HST optical data for many of the targets.  The sample galaxies are at distances ranging from 40 to 180~Mpc, are at various stages of merging, interaction, or isolation and are  also observed with optical spectroscopy at SALT/RSS in Sutherland, South Africa (e.g. \cite{vaisanen08}).  Local LIRGs might physically represent more closely high-redshift star-formation (SF), where a diversity of modes of strong SF are present (e.g. \cite{rodighiero11}). 

Our survey is dubbed SUperNovae and starBursts in the InfraReD, {\em SUNBIRD}, for the twofold science aim:  We search for very obscured core-collapse SNe close to the nuclei of star-forming galaxies to determine the total SFR in the local universe (see e.g. \cite{mattila12,kankare12}), and study the LIRGs themselves to have a uniform sample of local LIRGs to investigate the physical details of star-formation and its triggering and history as a function of interaction stage and type, environment, and metallicity (\cite{vaisanen08,vaisanen12}).   In these proceedings we report on the galaxies themselves, and in particular on their Super Star Cluster (SSC) populations which likely trace extreme forms of SF in the systems.

\section{Luminosity Functions of SSCs in the NIR}
\label{sec:2}

SSCs are ubiquitous in the AO images of our LIRGs, though the characteristics of the populations have intriguing differences.  As a first step of the SSC characterisation we have derived their luminosity functions in the $K$-band.  The analysis of the whole sample is still on-going, but we have fitted the LF slopes in a  $\Phi(L) dL \propto L^{-\alpha} dL$ distribution for a sub-sample of 10 galaxies (\cite{zara13a}).  Using NIR wavelengths we are able to probe SSC populations in obscured environments decreasing scatter which would plague optical observations in complex dusty environments.  This work also introduces SSC studies in hosts with higher SFRs than most previous studies.  Note that the SSC studies in the SUNBIRD sample concern the brightest clusters, those at $M_K  < -14$ mag, or typical masses of $>10^5 \, M_{\odot}$.

We find the LFs of our clusters reasonably well-fitted by a single power law with values of the index $\alpha$ ranging between 1.5 and 2.4 with a combined value of $\alpha \sim 1.8$ (Fig.~1) in the pilot sample. Interestingly, this value is less steep than a typical $\alpha \approx 2.2$ found in normal spiral galaxies (e.g. \cite{gieles06}).  Relatively shallow slopes for SSC LFs have also been found in blue compact dwarfs (\cite{adamo11}).  Correlations of the $\alpha$ parameter with SFR and other host galaxy properties are being investigated with the full sample.

Due to the host galaxy distances involved (median $D_L \sim 70$ Mpc), blending effects have to be taken into account. We investigated the effect of blending on both the photometry and LF shapes using Monte Carlo simulations, as well as by repeating photometry and LF fits for an artificially redshifted Antennae galaxy system (i.e. converting the HST image from 20 Mpc distance to resolutions at different distances as observed with different instruments).   
While blending tends to flatten LFs, our analyses show that $\delta \alpha$ is less than $\sim$0.1 in the luminosity range considered.  The simulations also show that the extracted SSC luminosities are generally dominated by a single dominant stellar cluster rather than several knots of star formation (Fig.~1bc).  Hence, while the apertures where SSCs are measured {\em are} large, ranging typically from 20 to 60 pc within the sample, we have reason to believe that in the bright and massive SSC range the results are representative of individual massive SSCs, rather than large complexes of clusters (though on an individual level such cases no doubt will also enter the sample).

\begin{figure*}
  \centering
  \includegraphics[width=  11.8 cm]{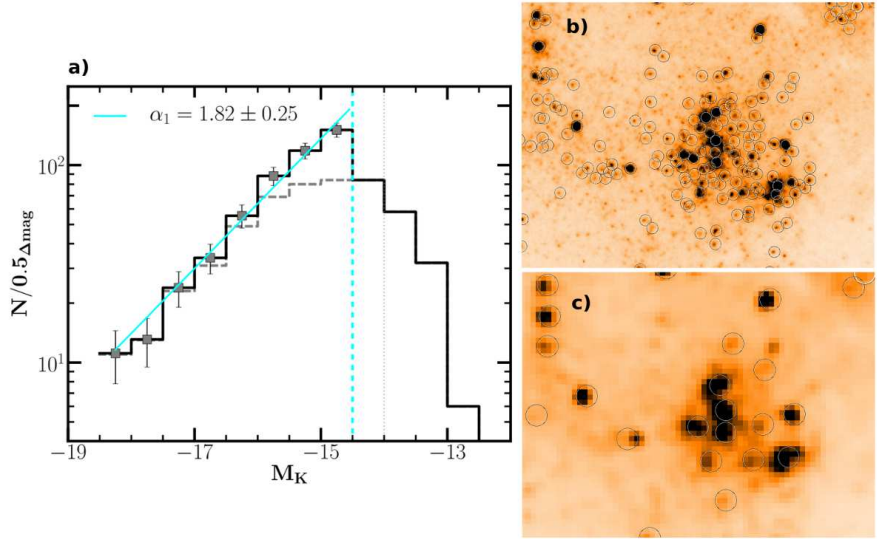}
  \caption{{\em Left:} A combined $K$-band SSC LF from 10 LIRGs, \cite{zara13a}.  A typical LF slope for the sample is $\alpha \approx 1.8$, less steep than for normal spiral galaxies. {\em Top right:} Region 'E' inside the Antennae system from \cite{whitmore10} $I$-band HST image. The {\em bottom right} shows the same as it would be observed if moved 4-times further, to 80 Mpc.  Only a fraction of the SSCs are detected anymore, though the properties of the brightest SSCs remain fairly intact.}
  \label{fig:fig1}
\end{figure*}


\section{Magnitude of the brightest cluster vs.\  SFR}
\label{sec:3}

Many studies have shown that there is an empirical relation between the $V$-band luminosity of the brightest SSC and the SFR of the host galaxy (e.g., \cite{larsen02, weidner04, bastian08}). Various reasons have been suggested to explain the relation, all highlighting the role star clusters play in understanding their host galaxy properties, and {\em vice versa}.  Using the AO data of the {\em SUNBIRD} Survey we have for the first time derived the relation in the NIR (\cite{zara13b}), and at the same time pushed the relation to significantly higher SFRs than reached before in statistical samples of host galaxies.

Fig.\ 2. shows our sample which is well fit by the relation $M_K \sim -2.6 \log$ SFR.   Again, we tested for distance-related effects, such as more distant, higher-SFR hosts having blended SSCs and hence artificially brighter SSCs: this was ruled out, though we exclude the three most distant galaxies at $>150$~Mpc.  The brightest cluster vs.\ SFR relation is similar to that found in the optical, though it appears somewhat steeper, as seen in the right panel of Fig.~2.  Since a comparison of an optical and NIR relation requires a color transformation, the most simple explanation for the slope difference would be a systematic color difference of the brightest SSC as a function of host SFR.  This could be an age effect, or perhaps most likely an extinction effect: the LIRGs are dustier than normal spirals and more extinction at high SFR would explain the difference in the respective slopes of the relation.

Physically even more interesting is the surprisingly tight relation of our NIR relation. The intuitive size-of-sample argument for the SFR vs. $M_{brightest}$  relation states that SSC populations at higher SFRs are larger, hence it is more likely to detect brighter SSCs from a random sampling of the LF.  We ran MC simulations of random SSC populations with given LF slopes and 'observing' the brightest ones.  In case of LFs in the normal range $\alpha = 1.5$ to 2.5, {\em and no upper limit to the LF}, the expected scatter is $\sim 1$ mag.  The scatter in the left panel of Fig.~2 is $\sigma \approx 0.6$ mag.  A natural explanation producing such smaller scatter would e.g.\ be a physical truncation in the LFs at the bright end.

\begin{figure*}
  \centering
  \includegraphics[width=  12.2 cm]{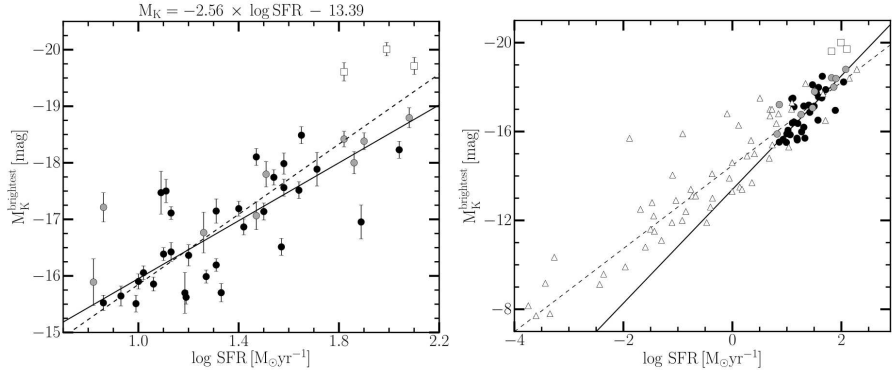}
  \caption{{\em Left:} An empirical relation between the $K$-band magnitude of the brightest cluster and the SFR of the galaxy. The solid line is our adopted weighted linear fit using the $D_L \leq 150\,\rm Mpc$ targets shown as circles. The open squares show galaxies further than this limit, with blending likely affecting the brightest cluster magnitude; the dashed line is a fit to all the data.  The relatively small scatter of the points in the diagram suggests an underlying physical cause, and not just a size of sample effect for the relation. {\em Right}: Equivalent data from the optical (as compiled by \cite{adamo11}) using a constant $V-K=2$ conversion are shown as triangles.  The solid line is the adopted best fit for our data, while the dashed line shows a slope of $\approx 2$ fitting the optical lower SFR data better, instead of the extrapolation of our $\approx 2.6$ slope.}
  \label{fig:fig3}
\end{figure*}

\section{Ages and masses of the LIRG SSCs and histories of hosts}

We aim to use the distribution and characteristics of SSCs to trace the history of violent SF in the host galaxies.  Many  of the LIRGs are members in interactions at various stages, and those that apparently are isolated often have complex nuclear structures, and/or evidence of extended ionised gas far outside their stellar content. The SSC distributions, if age-dated, may help in disentangling the roles played in their SFH by past major interactions, and/or minor mergers or secular processes.  Age and mass characteristics of the SSCs must be obtained with high confidence for this purpose.  In our {\em SUNBIRD} sample, in addition to the $K$-band, approximately a quarter of the galaxies have HST $B$ and $I$-band, some also have $V, J$, or $H$-band imaging available for SED fitting based on 3 or 4 available filters.  We use both the Starburst99 \cite{leitherer99} and the Yggdrasil \cite{zack11} models to do mass and age modelling.  To help in breaking the various degeneracies we constrain the extinctions and metallicities allowed based on the optical spectra we have obtained from SALT/RSS.  While the analysis is ongoing, we highlight some preliminary results  from one of the targets, IRAS~18293-3413, which has $BIHK$ photometry available. 

Some 500 SSCs are detected in IRAS~18293-3413, a LIRG with $log (L_{IR}) \sim 11.7$ consisting of a complex spiral with a small early type companion \cite{vaisanen08b}.  Approximately 300 SSCs with good quality data in all 4 bands are modelled using theYggdrasil models with Solar metallicity and extinctions limited to the range $A_V=1$ to 5.  Half of the SSC candidates appear to b
e young, less than 10 Myr of age and, interestingly, they are concentrated towards the inner 1-2 kpc.  A significant population, nearly 1/4 of them, are older with ages of several hundred Myr and their spatial distribution is more spread out.  Fig.~3 presents the mass and age distributions of these SSCs.  The mass functions are fairly well described by a $dN / dM \propto M^{\beta}$ relation with a $\beta \approx -2$ index, as is often found for more quiescent spirals as well, for all the various age populations. Other functions can also be fit, however.  The distributions turn over at fainter levels, but this is likely due to incompleteness which has not been assessed rigorously yet.  The age distribution, fit as a $dN / dt \sim t^{\gamma}$ relation, 
is often used to study cluster disruption models, though cluster {\em formation}, with its further correlation to variations of SFR in the host, is intricately involved.  Our analyses incorporate kinematical and stellar population analyses from spectroscopy in an attempt to disentangle the formation and disruption of SSCs to find a consistent picture with the global galaxy properties.  This is still ongoing for our sample.  Here we merely note the following:  the significant amount of older SSCs, seen e.g.\ as the "delta peak" in the age distribution at $\sim 1$ Gyr of the right panel of Fig.~3, suggests a previous interaction episode where SSC production has been even higher than currently.  This galaxy has thus likely been a ULIRG some 0.5 to 1 Gyr ago. The preliminary stellar population fits to the SALT/RSS full optical spectra reveal a SF history consistent with this scenario.


\begin{figure*}
  \includegraphics[width=  12.4 cm]{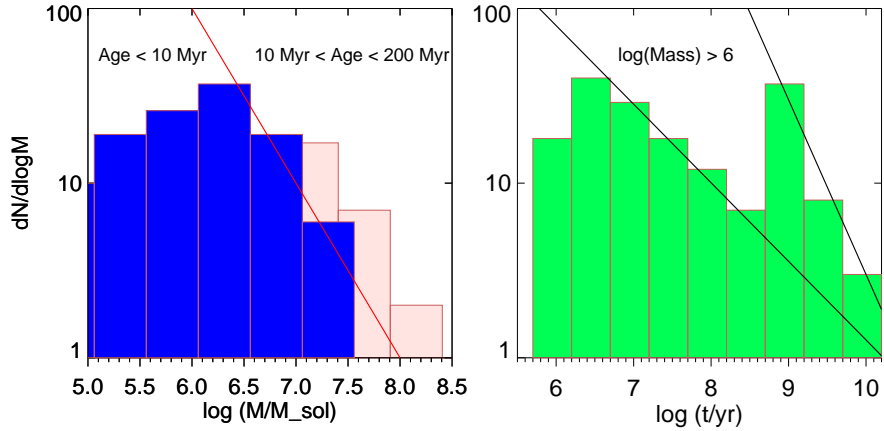}
  \caption{{\em Left:} Mass function of SSCs in IRAS~18293-3413.  The dark blue histogram shows the distribution of those younger than 10 Myr, while the light red distribution is for ones between the ages of 10 and 200 Myr. The oldest population is not shown here for clarity, but all distributions are reasonably well fit by a power law slope $\beta \sim -2$ displayed by the straight line. 
 {\em Right:}  The age distribution of SSCs, selected above $10^6 M_{\odot}$ to be complete at all ages.  A power law slope of $\gamma \sim -0.4$ (shown) fits well the the young and intermediate ages.  The oldest bins can be fitted with $\gamma \sim -1.0$ (also shown).  However, another explanation could be that the whole distribution
has one shallow slope with just a "delta peak" of SSC formation at $\sim 1$ Gyr, suggesting a strong interaction episode at the time.}
  \label{fig:fig3}
\end{figure*}


%
%
%
%

%
%
\begin{acknowledgement}
We thank the organizers of GH2013 for a stimulating conference and for the opportunity to present our ongoing research on SSCs.

\end{acknowledgement}


\begin{thebibliography}{99.}%


\bibitem{adamo11} 
{Adamo} A., {{\"O}stlin} G., {Zackrisson} E. 2011, MNRAS, 417, 1904

\bibitem{bastian08}
{Bastian} N. 2008, MNRAS, 390, 759


\bibitem{gieles06} Gieles M., Larsen S.S., Bastian N., Stein I.T., 2006, A\&A, 450, 129

\bibitem{kankare12} Kankare E., Mattila S., Ryder S., et al., 2012, ApJ, 744, L19

\bibitem{larsen02}
{Larsen} S.S. 2002, AJ, 124, 1393


\bibitem{leitherer99} Leitherer C.. 1999, ApJS, 123, 3

\bibitem{mattila12} Mattila S., Dahlen T., Efstathiou A., et al.,  2012, ApJ, 756, 111

\bibitem{zara13a}Randriamanakoto Z., V\"ais\"anen P., Kankare E., Kotilainen J., Mattila S., Ryder S., MNRAS, 431, 554

\bibitem{zara13b}Randriamanakoto Z., Escala A., V\"ais\"anen P., Ryder S., Kankare E., Kotilainen J., Mattila S. ApJ, 775, L38

\bibitem{rodighiero11} Rodighiero, G., Daddi E., Baronchelli I., et al., 2011, ApJ, 739, L40

\bibitem{vaisanen08}  {V{\"a}is{\"a}nen} P., {Mattila} S., {Kniazev} A., {et~al.}  2008, MNRAS, 384, 886

\bibitem{vaisanen08b}  {V{\"a}is{\"a}nen} P., Ryder, S., {Mattila} S., Kotilainen J.,  2008, ApJ, 689, L37

\bibitem{vaisanen12} V\"ais\"anen P.,  Escala A.,  Kankare E., et al., 2012, Journal of Physics Conf.\ Series, 372, 1 (arXiv:1202.6236)

\bibitem{weidner04}{Weidner}, C., {Kroupa}, P., \& {Larsen}, S.~S. 2004, MNRAS, 350, 1503

\bibitem{whitmore10} Whitmore B.C., Chandar R., Schweizer F., et al., 2010, AJ, 140, 75

\bibitem{zack11} Zackrisson E., et al., 2011, ApJ, 740, 13 

\end{thebibliography}
\end{document}